\documentstyle[prd,aps,preprint,epsfig,floats,graphicx,amssymb]{revtex}
\tightenlines
\input epsf.tex
\def\DESepsf(#1 width #2){\epsfxsize=#2 \epsfbox{#1}}

\begin{document}

\draft

\preprint{\vbox{
\hbox{OSU-HEP-02-15}
}}

\title{{\Large\bf Two--loop Neutrino Mass Generation}\\[0.07in]
{\Large\bf  and its Experimental Consequences}}
\author{{\bf K.S. Babu} and {\bf C. Macesanu}}
\address{~\\
Department of Physics\\ Oklahoma State University\\  Stillwater, OK
74078, USA}
\maketitle

\begin{abstract}
If neutrino masses have a radiative origin, their smallness can be
naturally understood even when lepton number violation occurs near
the weak scale.  We analyze a specific model of this type wherein
the neutrino masses arise as two--loop radiative corrections.
We show that the model admits the near bimaximal mixing
pattern suggested by the current neutrino oscillation data. Unlike
the conventional seesaw models, these two--loop models can be
directly tested in lepton flavor violating decays
$\tau \rightarrow 3\mu$ and $\mu \rightarrow e+\gamma$ as well as
at colliders by the direct observation of charged scalars needed for
the mass generation.  It is shown that
consistency with the neutrino oscillation data requires that the
leptonic rare decays should be within reach of forthcoming experiments
and that the charged scalars are likely to be within reach of the LHC.
\end{abstract}

\vskip1.0in


\newcommand{\beq}[1] {\begin{equation}\label{#1} }
\newcommand{\eeq} {\end{equation} }

\newcommand{\bea}[1]{\begin{eqnarray}\label{#1} }
\newcommand{\eea}{\end{eqnarray}}

\newcommand{\eps}{\epsilon}
\newcommand{\om}{\omega}

\newcommand{\oee}{\omega_{e e}}
\newcommand{\oem}{\omega_{e \mu}}
\newcommand{\oet}{\omega_{e \tau}}
\newcommand{\omm}{\omega_{\mu \mu}}
\newcommand{\omt}{\omega_{\mu \tau}}
\newcommand{\ott}{\omega_{\tau \tau}}




\newpage

\section{Introduction}

Recent neutrino oscillation experiments, notably Super-Kamiokande
and SNO, have provided a wealth of information on neutrino
masses and mixing pattern.  First and foremost, the solar \cite{solar}
and the atmospheric
neutrino data \cite{atm} are incompatible with the hypothesis of zero neutrino mass.
The inferred values of neutrino masses are in the sub--eV range.
The mixing angle $\theta_{12}$ relevant for
solar neutrino oscillations appears to be large ($\tan^2\theta_{12} \sim
0.4$) \cite{bahcall}, while the one for
atmospheric neutrino oscillations $\theta_{23} \sim \pi/4$ is near maximal.
Reactor neutrino experiments \cite{chooz} constrain the third mixing angle
$\theta_{13}$ to be relatively small, $\theta_{13} \leq 0.16$.  It will be of
great interest to seek a theoretical understanding of these observations.

The seesaw mechanism \cite{seesaw} has been the most popular explanation
of the small neutrino masses.  It is elegant and simple, and relies
only on dimensional analysis for the required new physics.  Current
neutrino data points to a seesaw scale of $M_R \sim 10^{10}-10^{15}$ GeV,
where lepton number violation occurs through the Majorana masses of the
right--handed neutrinos.  With such a high scale, effects of lepton flavor
violation in processes other than neutrino oscillation itself
becomes extremely small.  For example, the branching ratio for the decay
$\mu \rightarrow e+\gamma$ is of order $10^{-50}$ within the seesaw
extension of the Standard Model.

An alternative to the seesaw mechanism which also explains the smallness
of neutrino masses naturally is the radiative mass generation mechanism
\cite{zee1,cheng,zee2,babu,hall,mohap,ma,rp,bl}.
In this approach, neutrino masses are zero at the tree--level, and are
induced only as finite radiative corrections.  These radiative corrections
are typically proportional to the square of the charged lepton (or quark)
masses divided by the scale of new physics $\Lambda$.  The neutrino
masses  can then be in the sub--eV range even for $\Lambda \sim $ TeV.  In this case,
lepton flavor violation in processes other than neutrino oscillations may
become experimentally accessible.

In this paper we propose to analyze in detail the experimental consequences
of a specific model wherein the neutrino masses arise as two--loop radiative
corrections \cite{zee2,babu}.  The magnitude of these induced masses are of order
$m_\nu \sim [(f^2h)/(16 \pi^2)^2] (m_\tau^2/\Lambda)$, where $f,~h$ are dimensionless
Yukawa couplings and $\Lambda$ is the scale of new physics.  For $f \sim h
\sim 0.1$ and $\Lambda \sim$ 1 TeV, the neutrino mass will be of order 0.1 eV,
which is compatible with the oscillation data.  Within this model, it is required
that $f,~h \geq 0.1$ and $\Lambda \leq 1$ TeV, or else the induced neutrino
mass will be too small to be relevant experimentally.  These limits,
along with the requirement of large solar and atmospheric neutrino
oscillation angles, imply that
lepton flavor violation in processes such as
$\tau \rightarrow 3\mu$ and $\mu \rightarrow e+\gamma$ cannot be suppressed
arbitrarily.  We shall see that these decays are within reach of the
next round of rare decay experiments.
Furthermore, the upper limit on $\Lambda$ implies that
it is very likely that the
new scalars predicted by the model for neutrino mass generation
will be accessible for direct
observation at the LHC and perhaps even at Run II of the Tevatron.

A variety of radiative neutrino mass models exist in the literature.
The Zee model of neutrino mass \cite{zee1} is a popular example.
In this model, neutrino masses arise as one--loop
radiative corrections.  This model appears to be incompatible with the large
mixing angle MSW solution for the solar neutrino problem as it predicts the
relevant mixing angle to to very close to $\pi/4$ \cite{koide}.  In any case,
one would expect
rare leptonic processes to be more suppressed in the Zee model compared
to the two--loop neutrino mass model \cite{zee2,babu}.  This is because the neutrino
masses in the Zee model, being of one--loop origin,
 are of order $m_\nu \sim [fg/(16 \pi^2)]
(m_\tau^2/\Lambda)$, and in order for it to explain the atmospheric data,
one must choose either $f \leq 10^{-4}$ if $\Lambda \leq $ TeV, or
 $\Lambda \geq 10^6$ GeV if $f \sim 1$.
This should be contrasted with the two--loop neutrino mass model
where the Yukawa couplings are needed to be order 0.1 and $\Lambda$
of order TeV.

Another class of neutrino mass models that has been widely studied
is  supersymmetric models with
explicit $R$--parity violation \cite{hall}.  In this case, it turns out that
only one  neutrino acquires a tree--level mass,  the other two
obtain masses as one--loop radiative corrections.  The phenomenology
of such models, especially when  $R$--parity violation is soft, arising
only  through bilinear terms, has been thoroughly investigated \cite{rp}.

In Ref. \cite{bl} a general effective operator approach for small Majorana
neutrino masses has been presented.  It follows from that analysis that
if lepton number violation resides only in the leptonic sector and does
not involve the quarks, then the only interesting model for the current
neutrino oscillation data that also predicts observable lepton flavor
violation signals is the two--loop model of Ref. \cite{zee2,babu}.  This
is the main reason for revisiting the two--loop model in the present paper.

The outline of the paper is as follows. In Section II we will
have a quick review of the two--loop  neutrino mass model.  In Section III
we analyze quantitatively the constraints on the model
arising from neutrino oscillation data.  Here we also show how the
model can accommodate the near bi--maximal mixing pattern preferred by
the current data.  In Section IV, we analyze the constraints arising
from rare lepton decays.  Here we present bounds on the masses and
couplings of the charged scalars present in the model.  Section V is devoted
to the collider implications of the model.  We provide our conclusions
in Section VI.

\section{Description of the Model}

We start by reviewing the essential features of the two--loop neutrino
mass model \cite{babu,zee2}.  The gauge group of the model
is the same as the Standard Model (SM).
In addition to the SM particles the model introduces two charged scalars
$h^+$ and $k^{++}$ which are both singlets of $SU(3)_C$ and $SU(2)_L$.
These scalars have Yukawa couplings to the leptons, which can be
parametrized in terms of two complex matrices $f$ and $h$:
\beq{eq1}
{\cal L}_{Y} = f_{ab}\left( \psi^{T i}_{aL} C \psi^j_{bL} \right)
\eps_{ij} h^+ \
+ \ h'_{ab} \left( l^T_{aR} C l_{bR} \right) k^{++} + \hbox{h.c.}
\eeq
Here $ \psi_L$ stands for the left-handed lepton doublet, and $l_R$ for the
right-handed lepton singlet. $C$ is the charge conjugation matrix.  $i,j$ are
$SU(2)_L$ indices, while $a,b$ are generation indices.  The matrix $f$ is
antisymmetric ($f_{ab} = -f_{ba}$) due to Fermi statistics and antisymmetry
in the $SU(2)_L$ indices, while $h'$ is a symmetric matrix ($h'_{ab} =
h'_{ba}$).
The interaction Lagrangian in terms of the component fields
will then be
\beq{eq2}
{\cal L}_{Y} = 2 \left[
f_{e\mu} ( \bar{ \nu^c_e } \mu_L - \bar{ \nu^c_{\mu} } e_L ) \ + \
f_{e\tau} ( \bar{ \nu^c_e } \tau_L - \bar{ \nu^c_{\tau} } e_L ) \ + \
f_{\mu\tau} ( \bar{ \nu^c_{\mu} } \tau_L - \bar{ \nu^c_{\tau} } \mu_L ) \
\right] h^+
\eeq
$$
+ \left[ \
h_{ee} \bar{e^c} e_R + h_{\mu\mu} \bar{ \mu^c} \mu_R +
h_{\tau\tau} \bar{ \tau^c} \tau_R
 +  h_{e\mu} \bar{e^c} \mu_R +  h_{e\tau} \bar{e^c} \tau_R
 +  h_{\mu\tau} \bar{\mu^c} \tau_R
 \ \right] k^{++} + \hbox{h.c.}
$$
where we have defined $ h_{aa} = h'_{aa},  h_{ab} = 2 h'_{ab}$ for
$a \neq b$.

The Yukawa interactions of Eq. (1) conserves lepton number ($L$), as can be
seen by assigning two units of $L$ to the $h^+$ and $k^{++}$
fields.  The scalar piece of the Lagrangian contains a term
\beq{Lhk}
 {\cal L}_{h-k} = - \mu h^+ h^+ k^{--} + {\rm h.c.}
\eeq
which would then violate $L$.  In fact, one sees that the combination of
Eqs. (1) and (3) explicitly breaks lepton number.  This would lead to
the generation of Majorana neutrino masses at the two--loop level.  The relevant
diagram is shown in Fig. 1. The induced neutrino mass can be calculated to be

\beq{M2l}
({\cal M}_\nu)_{ab} = 8 \mu f_{ac} m_c h^*_{cd} m_d f_{db} I_{cd}
\eeq
where $m_c, m_d$ are the charged lepton masses
 and $ I_{cd} $ is the two loop integral function given by
\beq{Ieq}
I_{cd} = \int \frac{d^4 k}{(2\pi)^4} \int \frac{d^4 q}{(2\pi)^4} \
\frac{1}{(k^2 -m_c^2)} \frac{1}{(k^2 -m_h^2)}
\frac{1}{(q^2 -m_d^2)} \frac{1}{(q^2 -m_h^2)}
\frac{1}{(k-q)^2 -m_k^2} .
\eeq

\begin{figure}[t!] 
\centerline{
   \includegraphics[height=2.in]{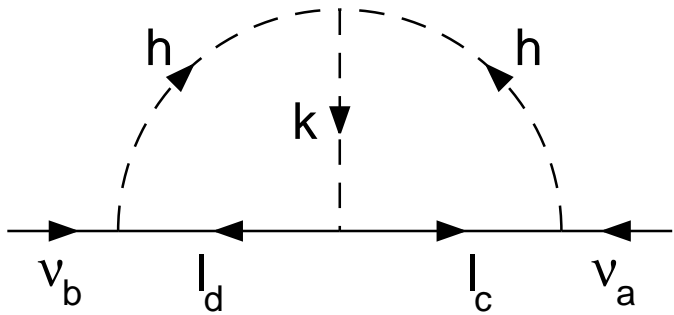}
   }
\caption{Diagram leading to finite neutrino mass at two loop.}
\label{nu_mass}
\end{figure}
\begin{figure}[hbt!] 
\centerline{
   \includegraphics[height=2.5in]{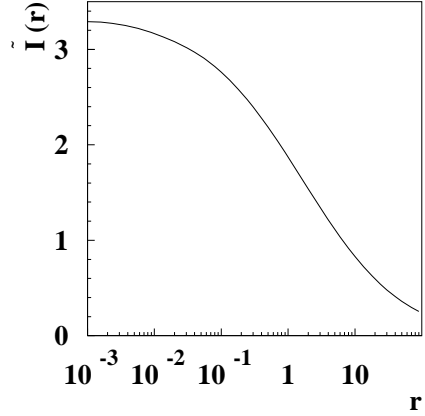}
   }
\caption{The dimensionless function $\tilde{I}(r)$
appearing in the evaluation of the two-loop mass integral.}
\label{mass_int}
\end{figure}

In order to evaluate the above integral, we may neglect the lepton massess
in the denominator, since these massess are much smaller than the charged
scalar masses $m_h$ and $m_k$. Then
\beq{Ieq2}
I_{cd} \simeq I = \frac{1}{(16 \pi^2)^2} \ \frac{1}{m_h^2} \
\tilde{I}\left(\frac{m_k^2}{m_h^2}\right).
\eeq
The dimensionless quantity $\tilde{I}$ defined as
\beq{Itil1}
\tilde{I} (r) = - \int_0^1 d x \int_0^{1-x} d y
\frac{1-y}{x + (r-1)y + y^2} \ \hbox{log} \frac{y(1-y)}{x+ry}
\eeq
is plotted in Fig. \ref{mass_int} for $r = m_k^2/m_h^2 < 100$.
The asymptotic behavior for small $m_k$ and
$m_k \gg m_h$ is given by:
\beq{asbeh}
\tilde{I} (r) \ \rightarrow \
 \left\{
    \begin{array}{ll}
{\displaystyle \frac{\hbox{log}^2r + \pi^2/3 - 1}{r} }
    & \hbox{for} \ r \gg 1 \\
\pi^2/3 & \hbox{for} \ r \rightarrow 0
    \end{array}
\right.
\eeq

\begin{figure}[ht!] 
\centerline{
   \includegraphics[height=2.5in]{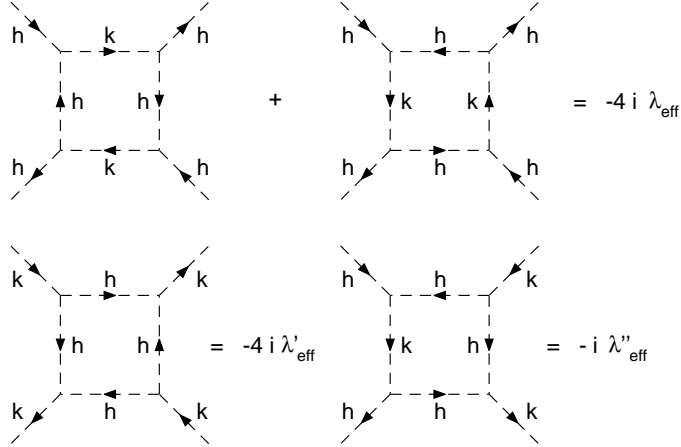}
   }
\caption{ Diagrams giving rise to the effective quartic interaction terms
for the $k$ and $h$ fields.}
\label{phi4}
\end{figure}

Before we estimate the induced neutrino masses and mixings, we wish to
make some remarks on the expected magnitude of the relevant coupling
parameters $f_{ab}, h_{ab}$ and $\mu$.  First, in order for perturbative
expansion to be sensible, the Yukawa couplings
$ f_{ab}, h_{ab}$ have to be of order unity or smaller.
For concreteness we shall take $f_{ab}< 1, ~h_{ab}<2$ (the $f_{ab}$ couplings
appear with a factor of 2 in Eq. (\ref{eq2})) (see further remarks at the end of
this section).
The dimensional parameter $\mu$ can be constrained as follows.
The Lagrangian terms in Eq. (\ref{Lhk}) lead to
an effective quartic interaction for the $h^+$ and $k^{++}$ fields
given by
\beq{Lphi4}
-{\cal L}_{eff } = \lambda_{eff} (h^+)^2 (h^-)^2 \ + \
\lambda'_{eff} (k^{++})^2 (k^{--})^2 \ + \
\lambda''_{eff} (h^+ h^-) (k^{--} k^{++})~.
\eeq
Evaluating the diagrams in Fig. \ref{phi4}, we obtain:
\bea{la1}
\lambda_{eff} & = & -\frac{1}{2 \pi^2} \frac{\mu^4}{(m_k^2 - m_h^2)^2}
\left[
 \frac{m_k^2 + m_h^2}{m_k^2 - m_h^2}
    \hbox{log} \left(\frac{m_k^2}{m_h^2}\right)
\ - \ 2 \right] \\
\lambda'_{eff} & = & -\frac{1}{4 \pi^2}\ \frac{\mu^4}{6 m_h^4} \nonumber \\
\lambda''_{eff} & = & -\frac{1}{\pi^2}\ \frac{\mu^4 m_k^2}{
    2(m_k^2-m_h^2)^3} \left[ \frac{m_k^2}{m_h^2} - \frac{m_h^2}{m_k^2}
    - 2 \hbox{log}\left( \frac{m_k^2}{m_h^2} \right) \right] \nonumber~.
\eea
In the limit where the two masses are almost equal ($m_k \approx m_h $)
$$
2 \lambda_{eff} \ \approx \ \lambda''_{eff} \ \approx \ 4\lambda'_{eff}
 \ = \ -\frac{1}{\pi^2}\ \frac{\mu^4}{6 m_h^4} .
$$

Since the effective couplings are negative, it follows that there must
exist in the tree--level Lagrangian terms similar to those in Eq.
(\ref{Lphi4}) whose couplings (call them $\lambda, \lambda', \lambda''$)
should be positive and greater in absolute value than
$\lambda_{eff}, \lambda'_{eff}, \lambda''_{eff}$ (otherwise the vacuum will be
unstable). Requiring that the theory be in the perturbative regime
with respect to these interactions ($\lambda, \lambda',
\lambda'' < 1$)
imposes the following constraints on the $\mu$ parameter:
\beq{mu_ctr}
\mu  \ < \  \left\{
\begin{array}{ll}
 m_h \times ( 6 \pi^2)^{1/4} & \hbox{if} \ m_k \approx m_h \\
 m_h \times ( 2 \pi^2)^{1/4} & \hbox{if} \ m_k \ll m_h \\
 m_h \times ( 24 \pi^2)^{1/4} & \hbox{if} \ m_k \gg m_h~.
\end{array}
\right.
\eeq

The upper limits on the Yukawa couplings $f_{ab}$ and $h_{ab}$ can be
made more precise by resorting to the renormalization group
equations (RGE) and by demanding that the couplings remain perturbative
to a momentum scale at least an order of magnitude higher than the weak
scale.  For example, consider the RGE evolution of the coupling $f_{\mu\tau}$:
$df_{\mu\tau}/dt = [3/(4 \pi^2)]f_{\mu\tau}^3 + ...$ $(t\equiv {\rm ln}\mu$).
From this it
follows that an initial value of $f_{\mu\tau} = 1$ will increase to
about 1.5 at a scale an order of magnitude higher.  Any larger initial
value of $f_{\mu\tau}$ will signal non-perturbative effects.  Similarly,
the coupling $\lambda_{eff}$ of Eq. (9) will receive a correction
proportional to $f_{\mu\tau}^4$, obtained from the RGE
$d\lambda/dt = 2f_{\mu\tau}^4/\pi^2
+ ...$.  An initial value of $f_{\mu\tau} = 1.5$ will result in $\lambda_{eff}
= 2.4$ at a momentum scale an order of magnitude higher, which will not be
in the perturbative regime.  An initial value of $f_{\mu\tau} = 1$ will
lead to $\lambda_{eff} = 0.47$ at a scale ten times higher, which is within
the perturbation theory.  These arguments suggest that $f_{ab}$ should be
smaller than about 1 for perturbation theory to be valid.  Similar arguments
suggest that $h_{ab} \leq 2$.


\section{Fitting the Neutrino Oscillation Data}

In this section, we first show how the two--loop neutrino mass model
fits the current neutrino oscillation data consistently and then derive
the constraints imposed by the data on the model parameters.

We use the standard parametrization of the neutrino mixing matrix
(the MNS matrix) in terms of three angles and a phase
\cite{pdg}:

\beq{rotm}
U \ = \left( \begin{array}{ccc}
1 & 0 & 0 \\
0 & c_{23} & s_{23} \\
0 &-s_{23} & c_{23}  \end{array}
\right) \
\left( \begin{array}{ccc}
c_{13} & 0 & s_{13} e^{-i\delta} \\
 0 & 1 & 0 \\
-s_{13}e^{i\delta}& 0 & c_{13}  \end{array}
\right) \
\left( \begin{array}{ccc}
c_{12} & s_{12} & 0 \\
-s_{12}& c_{12} & 0 \\
 0 & 0 & 1 \\ \end{array}
\right) \
\eeq
where $c_{ij} = \hbox{cos} \theta_{ij}, s_{ij} = \hbox{sin} \theta_{ij}$,
and $\theta_{ij}$ is the mixing angle between the flavor
eigenstates labeled by indices $i$ and $j$.
The transformation
\beq{mdiag}
 U^{T} {\cal M}_\nu U = {\cal M}_{diag}
\eeq
diagonalizes the neutrino mass matrix of Eq. (\ref{M2l}).

Before going further, we shall make some remarks on the possible values of the
neutrino masses.  An interesting feature of the two--loop neutrino mass model
is that, owing to the antisymmetric nature of the coupling matrix $f$,
the determinant of the mass matrix ${\cal M}_\nu$ is zero
\cite{babu}.  Thus one of the neutrino mass eigenvalues will be zero
within the model.  Although this result does not prevail when three--loop
and higher order corrections are included, the induced mass at these higher
loops will be extremely small and can be
neglected for all practical purposes.

If one uses the standard parametrization of $U$ shown in Eq. (12),
the neutrino mass eigenvalues should be allowed to be complex in general.
Since an overall phase can be removed by field redefinitions, there are
two relative (Majorana) phases in the mass eigenvalues in general.
However, since one of the mass eigenvalues is zero in the model under
discussion, there is a single Majorana phase, the relative phase of
the two non--zero mass eigenvalues.  Note that with the choice of $U$
as in Eq. (12), the elements of the matrix $f$ cannot be made real.

Experimental results indicate at
least two mass hierarchies (we shall not consider the LSND experiment
here):
from the atmospheric oscillation data ($\nu_{\mu} \leftrightarrow \nu_{\tau}$),
$\Delta m^2_{atm} \simeq 2.5 \times 10^{-3} ~\hbox{eV}^2$ \cite{atm},
and from the
solar neutrinos oscillation
($\nu_e \leftrightarrow \nu_{\mu} \hbox{ or } \nu_{\tau}$)
$\Delta m^2_{LMA} \lesssim 10^{-4} ~\hbox{eV}^2$ (the Large Mixing Angle MSW
solution) or even smaller \cite{solar}.
As a consequence, there are two possibilities for the diagonal mass matrix
${\cal M}_{diag}$ in
Eq. (\ref{mdiag}): the hierarchical form
\beq{mhier}
{\cal M}_{diag} =  \hbox{diag}( 0, m, M) \ \hbox{ with } |m| \ll |M|
\eeq
or the inverted hierarchy form:
\beq{minv}
{\cal M}'_{diag\pm} =  \hbox{diag}( M, \pm M + m, 0) \ \hbox{ with } |m| \ll |M| \ .
\eeq
We have allowed two possible signs in the inverted hierarchy case.  From the
point of view of fitting the neutrino oscillation data alone, the $+$
sign will be identical to the hierarchical form (Eq. (14)) since the matrices
differ only by an identity matrix.  However, the implications for the model
parameters are different, and thus the predictions for rare processes will be
different.  Note also that the model does not admit the case of three--fold
degenerate neutrino spectrum.

The atmospheric oscillation data implies that $|M|$ is about $0.05$ eV.
This already can give some information on the magnitude
of the parameters of the model. From
Eq. (\ref{M2l}), we have
\beq{c1}
\frac{8\mu}{(16 \pi^2)^2}\ \frac{|f|^2 \ |\Omega|}{m_h^2} \
\tilde{I} \approx  \ |M|
\ \approx \ 5 \times 10^{-12}~ \hbox{GeV}
\eeq
where $|f| = \hbox{max}(f_{ab}), \
|\Omega| = \hbox{max}(\omega_{ab} = m_a h_{ab}^* m_b)$. Taking $\tilde{I}$
to be of order unity and $\mu$ to have its maximum value of about $3 m_h$,
the above
equation becomes
\beq{c2}
\frac{|f|^2 \ |\Omega|}{m_h} \ \approx  \ 0.5 \times 10^{-8} ~\hbox{GeV}~.
\eeq
Since $m_h$ has to be at least of order $100$ GeV (otherwise $h^+$
 would have already been
seen in collider experiments),  the
larger of the couplings $f_{ab},\
\omega_{ab}/\hbox{GeV}^2$ have to be at least  of  the order
$10^{-2}$.
This raises the possibility that lepton flavor violation will be observable
in rare decays.  There may even be potential conflict with the current
limits.  We shall discuss these constraints quantitatively in the next
section.  Our conclusion there is that presently there is no conflict with
the data, but the rare decays should be accessible to the next round
of experiments.

In the rest of this section we shall  perform a more detailed analysis of the
neutrino mixings in the two--loop neutrino mass model. Our first aim is
 to reproduce  the neutrino mass hierarchy and the three neutrino mixing angles.
 For the case of the atmospheric oscillations,
the SuperKamiokande data indicates that the mixing is nearly maximal:
$|\theta_{23}| \simeq \pi/4$.
On the other hand, the $\nu_e \leftrightarrow \nu_{\tau}$ mixing angle
is close to zero; the reactor neutrino experiments \cite{chooz}
have set a limit
$\sin \theta_{13} < 0.16$. For the solar neutrino oscillations,
the SNO and SuperKamiokande
results indicate that while the mixing is not exactly maximal,
the mixing angle $\theta_{12}$  is of order one, with a preferred value
$\hbox{tan}^2 \theta_{12} \approx 0.4$.

In a first approximation let us take $\theta_{23} =
\theta_{12} = \pi/4$ and $\theta_{13} = 0$.  This is the exact
bimaximal mixing limit.
Then the neutrino mass matrix in the
flavor basis should have the form
\beq{m_nu}
\hat{\cal M}_{\nu} = \ U\ {\cal M}_{diag} U^{T} = \frac{1}{2}\left(
\begin{array}{ccc}
m & m/\sqrt{2} & -m/\sqrt{2} \\
m/\sqrt{2}  & M + m/2 & M - m/2 \\
-m/\sqrt{2}  & M - m/2 & M + m/2  \end{array}
\right)
\eeq
for the hierarchical case, or
\beq{m_nup}
\hat{\cal M}'_{\nu+} = \ U\ {\cal M}'_{diag+} U^{T} = \frac{1}{2}\left(
\begin{array}{ccc}
2M + m & m/\sqrt{2} & -m/\sqrt{2} \\
m/\sqrt{2}  & M + m/2 & -M - m/2 \\
-m/\sqrt{2}  & -M - m/2 & M + m/2 \end{array}
\right)
\eeq
\beq{m_nup2}
\hat{\cal M}'_{\nu-} = \ U\ {\cal M}'_{diag-} U^{T} = \frac{1}{4}\left(
\begin{array}{ccc}
2 m & (-2M + m)/\sqrt{2} & (2M - m)/\sqrt{2} \\
(-2M + m)/\sqrt{2}  &  m &  -m \\
(2M - m)/\sqrt{2}  &  -m &  m  \end{array}
\right)
\eeq
when the diagonal mass matrix has an inverted hierarchy form.
The mass matrices above will get small corrections due to the
deviation of the mixing angles from the exact bimaximal limit.

On the other hand, by expanding Eq. (\ref{M2l}),
the neutrino mass matrix from the two loop model can be
written in the form
{\normalsize
\beq{M2lexp}
  {\cal M}_{\nu} = \zeta
  \times \left(
\begin{array}{ccc}
\eps^2 \om_{\tau \tau} + 2 \eps \eps' \om_{\mu \tau} + \eps'^2 \om_{\mu \mu}
 \ , &
\eps \om_{\tau \tau}  + \eps' \om_{\mu \tau} - \eps \eps' \om_{e \tau}
  &
-\eps \omt -\eps' \omm - \eps^2 \oet  \\
 &  \hbox{~~~~~~~~~} - \eps'^2 \om_{e \mu} \ , &
 \hbox{~~~~~~~~~} - \eps \eps' \oem \\
. & \ott -2 \eps' \oet + \eps'^2 \oee \ , &
-\omt -\eps \oet + \eps' \oem  \\
 & & \hbox{~~~~~~~~~} + \eps \eps' \oee \\
. & . & \omm + 2 \eps \oem + \eps^2 \oee
\end{array}
\right)
\eeq
}
where we have defined
\begin{eqnarray}
\zeta &\equiv& \frac{8\mu}{(16 \pi^2)^2}\ \frac{f_{\mu \tau}^2}{m_h^2}
   \ \tilde{I}, ~~
\om_{ab} \equiv m_a h_{ab}^* m_b, \nonumber \\
\eps &\equiv& f_{e \tau}/f_{\mu \tau} , ~~
\eps' \equiv f_{e \mu}/f_{\mu \tau}~.
\end{eqnarray}
(We have written explicitly just the elements
above the diagonal in Eq. (21),
since the matrix is symmetric). Let us assume for the moment
that there is not a strong hierarchy among the $f_{ab}, h_{ab}$
couplings; in this case, the hierarchy among the lepton masses
($m_e \ll m_{\mu}, m_{\tau}$) allows us to neglect the
$\eps \om_{ea}, \eps' \om_{ea}$
quantities with respect to $\omm, \omt, \ott$. With this
simplification, the neutrino mass matrix will become
\beq{M2lsimp}
{\cal M}_{\nu} \simeq \zeta \left(
\begin{array}{ccc}
\eps^2 \ott + 2 \eps \eps' \omt + \eps'^2 \omm
 \ , &
\eps \ott  + \eps' \omt   & -\eps \omt -\eps' \omm \\
. & \ott  \ , & -\omt  \\
. & . & \omm~.
\end{array}
\right)
\eeq

Comparing with Eq. (\ref{m_nu}), we see that a choice
\beq{app_sol1}
\omm \approx  - \omt \approx \ott  ,\ \ \eps \approx \eps' \approx 1
\eeq
gives a good fit to the hierarchical neutrino mass matrix
(the mass hierarchy in this case being $m/M$ of order $\Delta \om / \om$).
For the inverse mass hierarchy case, the choice
$ \omm \approx  \omt \approx \ott $ would  fit
the lower right hand side $2 \times 2$ block of
the neutrino mass matrices in Eq. (\ref{m_nup}),
but the fact that there are large elements in the first row of these matrices
will require that there is a large hierarchy among the $f$ couplings; that
is, $\eps, \eps'$ should be of order $M/m$ or larger. In this case, though,
the simplification made to reach the form of Eq. (\ref{M2lsimp})
may not be entirely justified.

 The question arises then if it is possible to solve for the parameters
$\eps , \om$ exactly, without any approximation, in terms of the
elements of $\hat{\cal M}_\nu$.  Let's consider the general
equation
\beq{Meq1}
{\cal M}_{\nu} \equiv \frac{\zeta}{f_{\tau \mu}^2} \ f \om f^{T} = \hat{\cal M}
\eeq
and try to solve for the parameters
$\eps,  \eps' , \om_{ab}$ in terms of the $m_{ij}$ elements of the general
matrix $\hat{\cal M}$ (the only constraints on this matrix are that it be symmetric and
of zero determinant). At first glance, this task may seem impossible; there
are five independent equations, each of them of order three in the unknown
parameters. However, the particular form of the two loop mass matrix
${\cal M}_{\nu} \sim f \om f^{T}$, allows for considerable simplification of the
problem.

Note that $f$, having zero determinant (as a consequence of
antisymmetry),  has an eigenvector with zero as an eigenvalue:
\beq{v0}
v_0^{T} = ( 1 , -\eps , \eps' ) ; \ \ f v_0 = 0 \ .
\eeq
Moreover, $v_0$ is also an eigenvector of the ${\cal M}_{\nu}$ mass matrix;
multiplying
both sides of the matrix equation (\ref{Meq1}) by $v_0$ we obtain
\beq{v0eq}
\hat{\cal M} v_0 = 0 \ .
\eeq
The vector equation above allows solving for the parameters $\eps, \eps'$
in terms of the elements of the matrix $\hat{\cal M}$ (denoted by $m_{ij}$):
\beq{eps_sol}
\eps = \frac{m_{12} m_{33} - m_{13} m_{23}}{m_{22} m_{33} -m_{23}^2}, \
\eps' = \frac{m_{12} m_{23} - m_{13} m_{22}}{m_{22} m_{33} -m_{23}^2} \ .
\eeq
The surprising result is that the $\eps, \eps'$ parameters are completely
determined by the mass matrix elements $m_{ij}$, regardless of what the
$\om$ parameters might be.

Once $\eps$ and $\eps'$ are determined, there are six $\om$ parameters left,
and three independent equations.
These equations are linear in $\om$, and therefore easy to solve.
For example, we could use the 22, 23 and 33 matrix
equalities
\bea{om_eq}
\zeta \ ( \ott -2 \eps' \oet + \eps'^2 \oee ) & = & m_{22} \\
\zeta \ (  -\omt -\eps \oet + \eps' \oem  + \eps \eps' \oee ) & = & m_{23}
\nonumber \\
\zeta \ ( \omm + 2 \eps \oem + \eps^2 \oee ) & = & m_{33} \nonumber
\eea
to eliminate any three of the $\om$ parameters in terms of the other three.

In the following, we will apply this general analysis to the two
situations resulting from the  two possible hierarchies
(Eqs. (\ref{mhier},\ref{minv})) in the neutrino masses.

\vspace{0.2cm}
\noindent
$\bullet$ In the case of the hierarchical neutrino
mass matrix,  Eqs. (\ref{eps_sol}) read as
\beq{e_sol}
\eps \ = \ \hbox{tan} \theta_{12} \
\frac{\hbox{cos} \theta_{23}} {\hbox{cos} \theta_{13}} \ + \
\hbox{tan} \theta_{13} \ \hbox{sin} \theta_{23} \ e^{-i\delta}
\eeq
$$
\eps' \ = \ \hbox{tan} \theta_{12} \
\frac{\hbox{sin} \theta_{23}} {\hbox{cos} \theta_{13}} \ - \
\hbox{tan} \theta_{13} \ \hbox{cos} \theta_{23} \ e^{-i\delta}
$$
for  general mixing angles.
Taking $\theta_{23} \approx \pi/4, \theta_{13} \approx 0$, we have
\beq{e_sol2}
\eps \ \approx \ \eps' \ \approx \ \frac{\hbox{tan} \theta_{12}}{\sqrt{2}}.
\eeq
Then, for the LMA  and LOW solutions to the solar neutrino problem,
the $\eps$ and
$\eps'$ parameters are in the range $0.4 - 0.5$, with the difference between
them being of order $2 \theta_{13}$.

As mentioned above, for the case when $\eps, \eps'$ are of order unity (or
smaller)  the $\eps \om_{ea}$ terms  can be neglected with respect to
$\omm, \omt, \ott$.   Eq. (\ref{om_eq}) becomes
\bea{om_eq1}
\zeta \ \ott & = &  M/2  \ + \ m\ \hbox{cos} ^2 \theta_{12}
  (1 - \theta_{13} e^{i\delta} \hbox{tan} \theta_{12})/2  \\
\zeta \ \omt & = & -M/2 \ + \ \ m\ \hbox{cos} ^2 \theta_{12} /2 \nonumber \\
\zeta \ \omm & = &  M/2  \ + \ m\ \hbox{cos} ^2 \theta_{12}
  (1 + \theta_{13} e^{i\delta} \hbox{tan} \theta_{12})/2 \nonumber
\eea
in the $\theta_{23} = \pi/4$, small $\theta_{13}$ limit. The relative
difference between the magnitudes of the large $\om$ parameters is
$|\omm -\omt|/ |\omm| \simeq 2(m/M) \hbox{cos} ^2 \theta_{12} $, about
0.1 for the LMA solution, and 0.01 for the LOW solution.

The SMA solution to the solar neutrino oscillations,
although strongly disfavored by experimental data, can also be accommodated
in this model; in this case, $\eps, \eps'$ would be of order max($\theta_{12},
\theta_{13}) < 1/10$. The relations (\ref{om_eq1}) still hold.

Before going further, let us consider the reverse problem, that is, determining
the masses and mixing angles in terms of the $\eps$ and $\om$ parameters.
Eq. (\ref{e_sol}) can inverted for two angles in terms of a third:

\bea{an_sol}
 \hbox{tan} \theta_{13} \ & = &\ e^{i\delta}(\eps \ \hbox{sin} \theta_{23}
 \ - \ \eps' \ \hbox{cos} \theta_{23}) \\
\hbox{tan} \theta_{12} \ & = & \ \hbox{cos} \theta_{13}
(\eps \ \hbox{sin} \theta_{23}
 \ + \ \eps' \ \hbox{cos} \theta_{23} )  \ .\nonumber
\eea
Thus, we see that, provided that $\theta_{23} \approx \pi/4$, $\eps$ and $\eps'$
being almost equal implies that the mixing angle
$\theta_{13}$  is close to zero.

\vspace{0.2cm}
\noindent
$\bullet$ For the inverted mass hierarchy case,
the relations (\ref{eps_sol}) read as
\beq{esoli}
\eps = - \hbox{sin} \theta_{23}\ \hbox{cot} \theta_{13} \ e^{-i\delta} \ \ ,
\ \ \eps' =  \hbox{cos} \theta_{23}\ \hbox{cot} \theta_{13} \ e^{-i\delta} ~.
\eeq
Note that here $\theta_{13} = 0$
would require $f_{\mu \tau} = 0$. Barring this singular case,
we see that the parameters $\eps, \eps'$ are both of order $0.7/ \theta_{13}
\ \gtrsim \ 5 $. Then, neglecting the $\eps \ \omega_{ea}$ terms is not
necessarily well justified. While a general analysis is possible, for the
sake of simplicity we will restrict ourselves to the case when such terms
are small.

The equations determining the $h_{ab}$ coupling constants are:
\bea{om_eq2}
\zeta \ \ott & = &  M/2  \ + \ m\ \hbox{cos} ^2 \theta_{12}
  (1 - \theta_{13}  e^{i\delta} \hbox{tan} \theta_{12})/2  \\
\zeta \ \omt & = & M/2 \ + \ \ m\ \hbox{cos} ^2 \theta_{12} /2 \nonumber \\
\zeta \ \omm & = &  M/2  \ + \ m\ \hbox{cos} ^2 \theta_{12}
  (1 + \theta_{13} e^{i\delta} \hbox{tan} \theta_{12})/2 \nonumber
\eea
for the + sign in Eq. (\ref{minv}), and
\bea{om_eq3}
\zeta \ \ott & = &  (m\ \hbox{cos} ^2 \theta_{12}) /2  \ +
\ M\ \hbox{cos} 2\theta_{12} \
  (1 - 4\theta_{13} e^{i\delta} \hbox{tan} 2\theta_{12})/2  \\
\zeta \ \omt & = &  -(m\ \hbox{cos} ^2 \theta_{12}) /2 \ + \
  (M\ \hbox{cos} 2\theta_{12})/2 \nonumber \\
\zeta \ \omm & = &  (m\ \hbox{cos} ^2 \theta_{12}) /2  \ +
\ M\ \hbox{cos} 2\theta_{12} \
  (1 + 4\theta_{13} e^{i\delta} \hbox{tan} 2\theta_{12})/2 \nonumber
\eea
for the $-$ sign. In this latter case, since $m/M \simeq 10^{-2}$ for
the LMA solution (or smaller for the other solutions to the solar neutrino
oscillations), and the 1-2 mixing is not quite maximal
(cos$2 \theta_{12} \simeq 0.44$ for LMA again), the scale in the lower
right-hand corner of the neutrino mass matrix is set by the the terms
proportional to the large mass $M$. This means that the 2-3
family symmetry visible in the exact
bimaximal form of the neutrino mass matrix Eq. (\ref{m_nup2})
does not necessarily survive the small corrections in the mixing angles
required by experimental results. Also, the relation
(\ref{app_sol1}) between the $\omega$ parameters
is somewhat broken; we have
\beq{app_sol2}
\ott \ :\ \omt \ :\  \omm \ \simeq \ (1 - 4\theta_{13} \hbox{tan} 2\theta_{12})
\ : \ 1 \ : \ (1 + 4\theta_{13} \hbox{tan} 2\theta_{12}).
\eeq
However, even in this case the $h_{\mu \mu}$ coupling is the largest one.

In summary, we have found that both the hierarchical neutrino spectrum
and the inverted hierarchical spectrum with near bi-maximal mixings can
be accommodated in the two--loop neutrino mass model without difficulty.
The parameters of the model are then mostly determined.  We now turn to
other experimental signatures of the model.

\section{Experimental Constraints}

Besides neutrino masses and mixings, the Lagrangian in Eq. (\ref{eq1})
leads to non-standard lepton flavor violating processes. For example, at tree level,
the second
line in Eq. (\ref{eq2}) will allow lepton number violating decays,
such as $\mu^- \rightarrow e^+ e^- e^-$ and $\tau^- \rightarrow \mu^- \mu^+\mu^-$,
while the first line will
give extra contributions to the standard decay of the leptons.
At one loop level, the $f_{ab}$  and $h_{ab}$ couplings will contribute to the anomalous
magnetic momentum of the $e$ and the $\mu$, and will allow decays such as
$\mu \rightarrow e \gamma$.
 Experimental costraints
will therefore impose limits  on the parameters of
the model.

In the following, we analyze these processes in turn.

\vspace{0.2cm}
\noindent
$\bullet$  {\it Lepton family number violating $\mu $ and $\tau$ decays.}
\vspace*{0.2in}

The partial widths for these decay $l_a^- \rightarrow l_b^+ l_c^-l_d^-$ is
given by
\beq{lnv_dec}
\Gamma( l_a^- \rightarrow l_b^+ l_c^- l_d^-) =
\frac{1}{8} \frac{m_a^5}{192 \pi^3}
\left| \frac{h_{ab} h^*_{cd}}{m_k^2} \right|^2
\eeq
in the limit when the masses of the decay products are neglected with respect
to the mass of the decaying particle. The limits experimental constraints
\cite{pdg} set on the $h_{ab} h_{cd}$ combinations are summarized in Table 1.

\vspace{0.2in}
\noindent
$\bullet$ {\it Muonium-antimuonium oscillations.}
\vspace*{0.2in}

The last line in Table 1 is the limit arising from  muonium-antimuonium
conversion process $\mu^+ e^- \rightarrow \mu^- e^+$
\cite{muonium_exp}, mediated by $t$-channel
$k^{++}$ exchange. Using Fierz rearrangement, the amplitude for this
process can be written as
$$
 \frac{1}{2} \frac{i}{m_k^2} \ \bar{u}(\mu^+) \gamma^{\alpha} P_R u(e^+) \
 \bar{u}(\mu^-) \gamma_{\alpha} P_R u(e^-) \
( h_{e e} h^*_{\mu \mu} )
$$
giving rise to an effective Lagrangian coupling coefficient \cite{muonium_th}
$$ G_{M\bar{M}} = \frac{\sqrt{2}}{8} \ \frac{ h_{e e} h^*_{\mu \mu}}{m_k^2} \ .
$$

\begin{table}[!t]
\begin{center}
\begin{tabular}{c c c }
\hline
\hline
\hbox{Process} & \hbox{Exp. bound} & \hbox{Constraint} \\
\hline
$\mu^- \rightarrow e^+ e^- e^-$ & Br.  $< 1.\times 10^{-12}$ &
        $|h_{e \mu} h^*_{ee}| / m_k^2 < 3.3\times 10^{-11}
        \ \hbox{GeV}^{-2}$\\
$\tau^- \rightarrow e^+ e^- e^-$ & Br. $< 2.9\times 10^{-6}$ &
        $|h_{e \tau} h^*_{ee}| / m_k^2 < 1.3\times 10^{-7}
        \ \hbox{GeV}^{-2}$\\
$\tau^- \rightarrow e^+ \mu^- \mu^-$ & Br. $< 1.5\times 10^{-6}$ &
        $|h_{e \tau} h^*_{\mu \mu}| / m_k^2 < 0.9\times 10^{-7}
        \ \hbox{GeV}^{-2} $\\
$\tau^- \rightarrow e^+ e^- \mu^-$ & Br. $< 1.7\times 10^{-6}$ &
        $|h_{e \tau} h^*_{e \mu}| / m_k^2 < 1.\times 10^{-7}
        \ \hbox{GeV}^{-2}$\\
$\tau^- \rightarrow \mu^+ e^- e^-$ & Br. $< 1.5\times 10^{-6}$ &
        $|h_{\mu \tau} h^*_{ee}| / m_k^2 < 0.9\times 10^{-7}
        \ \hbox{GeV}^{-2}$\\
$\tau^- \rightarrow \mu^+ \mu^- \mu^-$ & Br. $< 1.9\times 10^{-6}$ &
        $|h_{\mu \tau} h^*_{\mu \mu}| / m_k^2 < 1.\times 10^{-7}
        \ \hbox{GeV}^{-2}$\\
$\tau^- \rightarrow \mu^+ e^- \mu^-$ & Br. $< 1.8\times 10^{-6}$ &
        $|h_{\mu \tau} h^*_{e \mu}| / m_k^2 < 1.\times 10^{-7}
        \ \hbox{GeV}^{-2}$\\
$\mu^+ e^- \rightarrow \mu^- e^+$ &  $G_{M\bar{M}} < 0.003\ G_F $ &
        $|h_{e e} h^*_{\mu \mu}| / m_k^2 < 4.\times 10^{-7}
        \ \hbox{GeV}^{-2}$\\
\hline
\hline
\end{tabular}
\end{center}
\label{table1}
\caption{Constraints from lepton family number violating processes.}
\end{table}

\vspace{0.2cm}
\noindent
$\bullet$ {\it Anomalous magnetic momenta and $\mu \rightarrow e \gamma$ --type
processes.}
\vspace*{0.2in}

At the one--loop level, both $h^+$ and $k^{++}$ exchange contribute
to the processes $l_a \rightarrow l_b \gamma$.
 For the case of same leptons in the initial and final states ($l_a \equiv l_b$)
this leads to change in
the anomalous magnetic moment of the  lepton:
\beq{amm}
\delta(g-2)_a \ = \ \frac{4 m_a^2}{96 \pi^2}\
\left[
 -\frac{ (f^{\dag}f)_{aa} }{ m_h^2}   \ + \
 \frac{ (h^{\dag}h)_{aa} }{ m_k^2}
\right]~.
\eeq
For different leptons in the initial and final states, the decay width
for $l_a \rightarrow l_b \gamma$ (with $m_a > m_b$) is:
\beq{meg_gam}
\Gamma(l_a \rightarrow l_b \gamma)
\ = \ 2\ \alpha\ m_a^3
\left( \frac{m_a}{96 \pi^2} \right)^2
\left[
\left( \frac{ (f^{\dag}f)_{ba} }{ m_h^2} \right)^2 +
\left( \frac{ (h^{\dag}h)_{ba} }{ m_k^2} \right)^2
\right]
\eeq
where $\alpha$ is the fine structure constant. The limits
experimental results on anomalous magnetic moments
\footnote{For the anomalous magnetic moment of the muon, the
bound on the non-SM contribution is obtained by adding the experimental
error in the measurement of $\delta(g-2)_{\mu}$ \cite{bnl_mu},
the theoretical error in evaluating the SM prediction for this
quantity
\cite{theor_mu}, {\it and} the difference between the experiment and theory
values of $\delta(g-2)_{\mu}$}
and the
$l_a \rightarrow l_b \gamma$ processes \cite{pdg}
impose on the $f$ and $h$ couplings are summarized in Table 2.

\begin{table}[!ht]
\begin{center}
\begin{tabular}{c c c }
\hline
\hline
& & \\
\hbox{Process} & \hbox{Exp. bound} & \hbox{Constraint} \\
& & \\
\hline
& & \\
$ e \rightarrow e \gamma$ & $\delta(g-2)_e < 8. \times 10^{-12}$ &
${ \displaystyle \frac{|f_{e\mu}|^2 +|f_{e\tau}|^2}{m_h^2} }+
{ \displaystyle \frac{|h_{ee}|^2+|h_{e\mu}|^2
+ |h_{e\tau}|^2}{m_k^2} }
< { \displaystyle \frac{ 7.6 \times 10^{-3} }{\hbox{GeV}^2} }$\\
& & \\
$ \mu \rightarrow \mu \gamma$ & $\delta(g-2)_{\mu} < 1.2 \times 10^{-8}$ &
${ \displaystyle \frac{|f_{e\mu}|^2 +|f_{\mu\tau}|^2}{m_h^2} }+
{ \displaystyle \frac{|h_{e\mu}|^2+|h_{\mu\mu}|^2
+ |h_{\mu\tau}|^2}{m_k^2} }
< { \displaystyle \frac{2.6 \times 10^{-5} }{\hbox{GeV}^2} }$\\
& & \\
$ \mu \rightarrow e \gamma$ & Br. $ < 1.2 \times 10^{-11}$ &
${ \displaystyle \frac{| f_{e\tau}^* f_{\mu \tau}|^2}{m_h^4} }+
{ \displaystyle \frac{| h_{ee}^* h_{e\mu} + h_{e\mu}^* h_{\mu \mu} +
h_{e\tau}^* h_{\mu\tau}|^2}{m_k^4} }
< { \displaystyle \frac{ 1.7 \times 10^{-17} }{\hbox{GeV}^4} }$ \\
& & \\
$ \tau \rightarrow e \gamma$ & Br. $ < 2.7 \times 10^{-6}$ &
${ \displaystyle \frac{| f_{e\mu}^* f_{\mu \tau}|^2}{m_h^4} }+
{\displaystyle \frac{| h_{ee}^* h_{e\tau} + h_{e\mu}^* h_{\mu \tau} +
h_{e\tau}^* h_{\tau\tau}|^2}{m_k^4} }
< { \displaystyle \frac{ 2. \times 10^{-11} }{\hbox{GeV}^4} }$ \\
& & \\
$ \tau \rightarrow \mu \gamma$ & Br. $ < 1.1 \times 10^{-6}$ &
${ \displaystyle \frac{| f_{e\mu}^* f_{e \tau}|^2}{m_h^4} }+
{ \displaystyle \frac{| h_{e\mu}^* h_{e\tau} + h_{\mu\mu}^* h_{\mu \tau} +
h_{\mu\tau}^* h_{\tau\tau}|^2}{m_k^4} }
< { \displaystyle \frac{8.2 \times 10^{-12} }{\hbox{GeV}^4} }$\\
& & \\
\hline
\hline
\end{tabular}
\end{center}
\label{table2}
\caption{Constraints from $l_a \rightarrow l_b \gamma$ type processes.}
\end{table}

\vspace{0.2cm}
\noindent
$\bullet$ {\it $l_a \rightarrow \nu_a l_b  \bar{\nu_b}$  decays.}
\vspace*{0.2in}

The exchange of a $h^+$ boson will contribute to the semileptonic decays
of the $\mu$ and $\tau$ leptons.
 Using Fierz rearrangement of spinors, the amplitude of the
$h^+$ exchange diagram for the process $l_a \rightarrow \nu_a l_b  \bar{\nu_b}$
can be shown to be proportional to the SM diagram:
\beq{mudec}
A_{h^+} = A_{SM} \times 4 \frac{|f_{ab}|^2}{m_h^2} \frac{M_W^2}{g^2 }~.
\eeq
Therefore, the angular distribution of the decay products is not
affected, while the total decay rate is. For the case of  $\mu$ decay,
this implies
the redefinition of the Fermi decay constant (which is extracted
from the muon lifetime measurements):
\beq{Gf}
G_{\mu} = G_F \left( 1\ + \ 4 \frac{|f_{e\mu}|^2}{m_h^2} \frac{M_W^2}{g^2 }
\right)^2
= 1.16639(1) \times 10^{-5} \hbox{GeV}^{-2}
\eeq
where $G_F$ refers to the SM Fermi constant, which is related
to other well--measured quantities in the Standard Model as (in the on-shell
scheme)
\beq{gfos}
G_F = \frac{\pi \alpha}{\sqrt{2} M_W^2 (1-M_W^2/M_Z^2) (1-\Delta r)} .
\eeq
Here $\Delta r$ encodes the effect of radiative corrections, which depends
on the top quark and  the Higgs boson masses. The redefinition of $G_F$
from Eq. (\ref{Gf}) can
be interpreted then as a redefinition of the $\Delta r$ parameter;
$ \Delta r \rightarrow \Delta r + \delta \Delta r$, with
\beq{del_r}
\delta \Delta r = -8 \frac{|f_{e\mu}|^2}{m_h^2} \frac{M_W^2}{g^2 }
\eeq
Analyses indicate that variations in $\Delta r$ of order 0.002 are acceptable
in the Standard Model \cite{del_r} .
 This would translate into a constraint on  $f_{e\mu}$
which can be found on the first line in Table 3.

It is interesting to note that the $h^+$ exchange {\it adds constructively}
 to the
muon decay rate.  The neutron and the nuclear beta decay rates, on the other
hand, are unchanged compared to the Standard Model.  The value of the CKM
matrix element $V_{ud}$ extracted from beta decay measurements will have
to be re-interpreted in the present model.  We find an {\it upward shift}
in $V_{ud}$ compared to the SM by an amount given by $|\delta \Delta r|$
of Eq. (44).  Currently there is a 2.2 sigma anomaly in the unitarity
of the first row of the CKM matris:  $|V_{ud}|^2 + |V_{us}|^2+|V_{ub}|^2$
is smaller than by about 2.2 sigma \cite{pdg}.
With $|\delta \Delta r| = 0.002$, the
upward shift in the value of $V_{ud}$ can nicely reconcile this anomaly
within this model.

Table 3 also contains constraints on the $f$ couplings coming from
the widths of semileptonic $\tau$ decays \cite{pdg}(lines 2,3).
Since the redefined Fermi constant
is used to compute these decays, the constraints are on the differences
between the $f$ couplings.
Line 4 contains the constraint coming from $e-\mu$ universality
in $\tau $ decay \cite{em_univ}.

\begin{table}[!ht]
\begin{center}
\begin{tabular}{c c c }
\hline
\hline
& & \\
\hbox{Process} & \hbox{Exp. bound} & \hbox{Constraint} \\
& & \\
\hline
& & \\
$\mu^- \rightarrow \nu_{\mu} e^- \bar{\nu_e}$ & $|\delta \Delta r| < 0.002$ &
        $|f_{e \mu}|^2 / m_h^2 < 1.6\times 10^{-8} \ \hbox{GeV}^{-2}$\\
& & \\
$\tau^- \rightarrow \nu_{\tau} e^- \bar{\nu_e}$ & Br = 17.83 $ \pm $ 0.06 \% &
 ${ \displaystyle \left| \frac{|f_{e\tau}|^2 - |f_{e \mu}|^2}{m_h^2} \right| }
 < 3.4 \times 10^{-8} \ \hbox{GeV}^{-2} $ \\
& & \\
$\tau^- \rightarrow \nu_{\tau} \mu^- \bar{\nu_{\mu}}$ &
     Br = 17.37 $ \pm $ 0.07 \% &
 ${ \displaystyle \left| \frac{|f_{\mu \tau}|^2 -
 |f_{e \mu}|^2}{m_h^2} \right| }
 < 4. \times 10^{-8} \ \hbox{GeV}^{-2} $ \\
& & \\
$ e / \mu$ universality
 & ${ \displaystyle \frac{G_{\tau e}}{G_{\tau \mu}} } = 0.999 \pm 0.003 $ &
${ \displaystyle \left| \frac{|f_{e\tau}|^2 - |f_{\mu \tau}|^2}{m_h^2} \right| }
 < 2.5 \times 10^{-8} \ \hbox{GeV}^{-2} $ \\
& & \\
\hline
\hline
\end{tabular}
\end{center}
\label{table3}
\caption{Constraints from  $\mu$ decay and semileptonic $\tau$ decays.}
\end{table}

\noindent $\bullet$ {\it Neutrinoless double beta decay.}
\vspace*{0.2in}

From the approximate forms of the neutrino mass matrix given in
Eqs. (18)-(20), it follows that the effective neutrino mass relevant for
neutrinoless double beta decay is approximately $m(\beta\beta0\nu) \sim
m/2 \sim
10^{-3}$ eV in the hierarchical case as well as in the inverted
hierarchical case of Eq. (20).  This will be difficult to observe
in the near future.  On the other hand, in the inverted mass hierarchy
of Eq. (19), the effective mass is $m(\beta\beta0\nu) \simeq M \simeq 0.05$
eV, which should be observable in the next round of experiments.

\vspace{0.2cm}
Now, let us look at these constraints in light of the relations among the
$f_{ab}$ and $h_{ab}$ couplings imposed by neutrino masses and mixings results of
Sec. III.
Consider first the $f$ parameters. In the case of hierarchical form for
the neutrino mass matrix, from Eqs. (\ref{e_sol}), (\ref{e_sol2}) we have
\beq{f_sol}
f_{e\mu} \approx f_{e\tau} \approx \frac{f_{\mu \tau}}{2}~.
\eeq
The strongest constraint on the $f$ couplings comes then from the
$\mu \rightarrow e \gamma $ process.  From Table 2 we have
\beq{f_cons}
 \frac{f_{\mu \tau}^2}{2 m_h^2} \ < \ 0.4 \times 10^{-8} \ \hbox{GeV}^{-2}~.
\eeq
On the other hand, from the last of Eq. (\ref{om_eq1}), we have
\beq{massm_eq}
\frac{8\mu}{(16 \pi^2)^2} \ \frac{f_{\mu \tau}^2}{m_h^2} \ \tilde{I} \
m_{\mu}^2 h_{\mu\mu} \ \simeq \  \frac{M}{2}
\eeq
with $M \simeq \sqrt{2.5 \times 10^{-3}} \ \hbox{eV} =
5 \times 10^{-11} \hbox{GeV}$ (here we have
neglected the contribution of
the terms proportional to $m_e$).
Plugging in the numbers we find
\beq{mh1eq}
m_h \ \simeq \ 10^5 \ \left( \frac{\mu}{m_h} \right)
\ f_{\mu \tau}^2 h_{\mu \mu} \
\tilde{I} \ \hbox{GeV}~.
\eeq
Assume for now that $m_k$ is smaller or of the same order as $m_h$; then
$\tilde{I} \approx 2$.
Using the upper limits on the $h$ couplings and $\mu/m_h$ ratio derived in
Sect. II, we get the following allowable range for the $h^+$ scalar mass:
\beq{mh2eq}
10^4\ f_{\mu \tau} \ \hbox{GeV} \ < \ m_h \ < \
10^6\ f_{\mu \tau}^2 \ \hbox{GeV}~.
\eeq
Here  the lower limit comes from the $\mu \rightarrow e \gamma$ constraint
Eq. (\ref{f_cons}), while the upper limit comes from the
neutrino mass equations above. We see then that when the $ f_{\mu \tau}$
coupling takes its maximum value (of order unity), the $h^+$ scalar will be
out of
reach of future colliders. However, this is the most unfavorable case; if
the $h, f$, and $\mu$ couplings take values smaller than the highest values
admissible, the upper bound on $m_h$ moves to lower values. For example,
if $ f_{\mu \tau} \simeq 0.1$, $m_h$ has to be smaller than 10 TeV.
Note also that Eqs. (\ref{f_cons}), (\ref{mh1eq}) impose a lower limit on the
strength of the $ f_{\mu \tau}, h_{\mu \mu}$ couplings:
\bea{lowl}
 f_{\mu \tau} & \ > \ & \frac{0.1}{h_{\mu \mu} \ \tilde{I} }\
 \left( \frac{\mu}{m_h} \right)^{-1} \ \gtrsim \ 1. \times 10^{-2} \\
 h_{\mu \mu} & \ > \ & \frac{0.1}{f_{\mu \tau} \ \tilde{I} }\
 \left( \frac{\mu}{m_h} \right)^{-1} \ \gtrsim \ 1.7 \times 10^{-2}~.
\nonumber
\eea
This requires that $m_h$ has to be greater than about 100 GeV. Also, the
lower bound on $ h_{\mu \mu}$ justifies neglecting the $\oee, \oem$
terms in the neutrino mass matrix; indeed, for the $\oem$ term to
be of the same order of magnitude as $\omm$, we would need
$h_{e \mu} \approx  200 \times h_{\mu \mu}$ which will move the theory
 into the nonperturbative regime.

We conclude from the preceding discussions that the decay $\mu \rightarrow
e+\gamma$ should be accessible to the next round of rare decay
experiments.  Some of the model parameters are already excluded by
current limits.  For example, if $m_h \sim 1$ TeV,  any value of
$f_{\mu\tau} \geq 2 \times 10^{-3}$ will be inconsistent with current
limits.  There are plans to improve the present limit on $\mu \rightarrow
e+\gamma$ by several orders of magnitude in the near future \cite{kuno}.
This decay is predicted to be within reach of these improved experiments.


The leptonic phenomenology
constraints on the $h_{ab}$ coupling constants and  the mass of the
doubly charged scalar  $m_k$ are somewhat weaker. All constraints
from processes involving the electron can be made to go away by choosing
$h_{ee}, h_{e \mu}$ and/or $h_{e \tau}$ to be close to zero (which
is allowed by the neutrino masses and mixings pattern). Then, the strongest
constraint on the  $h_{ab}/m_k$ ratio comes from the
$\tau^- \rightarrow \mu^+ \mu^- \mu^-$ process; using Eq.
(\ref{app_sol1}) to relate  $h_{\mu \tau}$ and $h_{\mu \mu}$,
this  constraint reads as
$$
\frac{h_{\mu \mu}}{m_k} \ < \ 1.3 \times 10^{-3}\ \hbox{GeV}^{-1}
$$
which can be easily satisfied without violating the other constraints discussed
previously. Note also that it is not required that $h_{ee}, h_{e \mu}$ and
$h_{e \tau}$ be close to zero; independent constrains on these quantities
from Table 1 and Table 2 are:
$$
\frac{| h_{ee} h_{\mu \mu}^* |}{m_k^2} < \frac{4. \times 10^{-7}}{ \hbox{GeV}^2},
\
\frac{| h_{e \mu} h_{\mu \mu}^* |}{m_k^2} < \frac{4. \times 10^{-9}}
{ \hbox{GeV}^2}, \
\frac{| h_{e \tau} h_{\mu \mu}^* |}{m_k^2} < \frac{9. \times 10^{-8}}
{ \hbox{GeV}^2},
$$
allowing any of them to be of the same order of magnitude as $h_{\mu \mu}$
(although not both $h_{e e}$ and $h_{e \mu}$ at the same time).

The bounds on the parameters of this model
are more stringent for the inverted mass hierarchy case. In this case
$ |f_{e \mu}| \simeq |f_{e \tau}| \simeq |f_{\mu \tau}|/ \hbox{sin}\theta_{13}$
are the larger $f$ couplings. In terms of $f_{e \mu}$ the bounds
on $m_h$ can be written:
\beq{mh3eq}
2 \sqrt{\hbox{sin}\theta_{13}} \times 10^4\ f_{e \mu} \ \hbox{GeV}
\ < \ m_h \  \simeq \
2 \hbox{sin}^2\theta_{13} \times 10^5 \ \left( \frac{\mu}{m_h} \right)
\ f_{e \mu }^2 h_{\mu \mu} \
\tilde{I} \ \hbox{GeV}
\eeq
$$
\lesssim \ 2 \hbox{sin}^2\theta_{13} \times 10^6 \ f_{e \mu }^2 \ \hbox{GeV}
$$
with $f_{e \mu} < 1$. The first inequality comes from the
$\mu \rightarrow e \gamma $ process in Table II, while the
last one comes from the neutrino mass equations (\ref{om_eq2})
(we considered the + case here).
Note that the allowed interval is narrower and shifted toward lower
values of $m_h$. The lower bounds on the $f_{e\mu}$ and and
$h_{\mu \mu}$ in Eqs. (\ref{lowl}) are increased by a factor
$1/(\hbox{sin}\theta_{13})^{3/2}$. This also imposes a lower
bound on the value of the $\theta_{13}$ mixing angle:
\beq{sint}
\hbox{sin}\theta_{13} \ > \ 0.046~
\eeq
which should be testable in neutrino oscillation experiments.
These constraints are relaxed by a factor $ \hbox{cos}2\theta_{12}
(1+ 4 \theta_{13} \hbox{tan} 2\theta_{12})$
$(\simeq 0.44(1+ 8 \theta_{13})$ for $\hbox{tan}^2 \theta_{12} = 0.4$)
for the $-$ sign in the inverted mass hierarchy case (using
Eqs. (\ref{om_eq3})
for the neutrino mass matrix elements).

\begin{figure}[t!] 
\centerline{
   \includegraphics[height=3.5in]{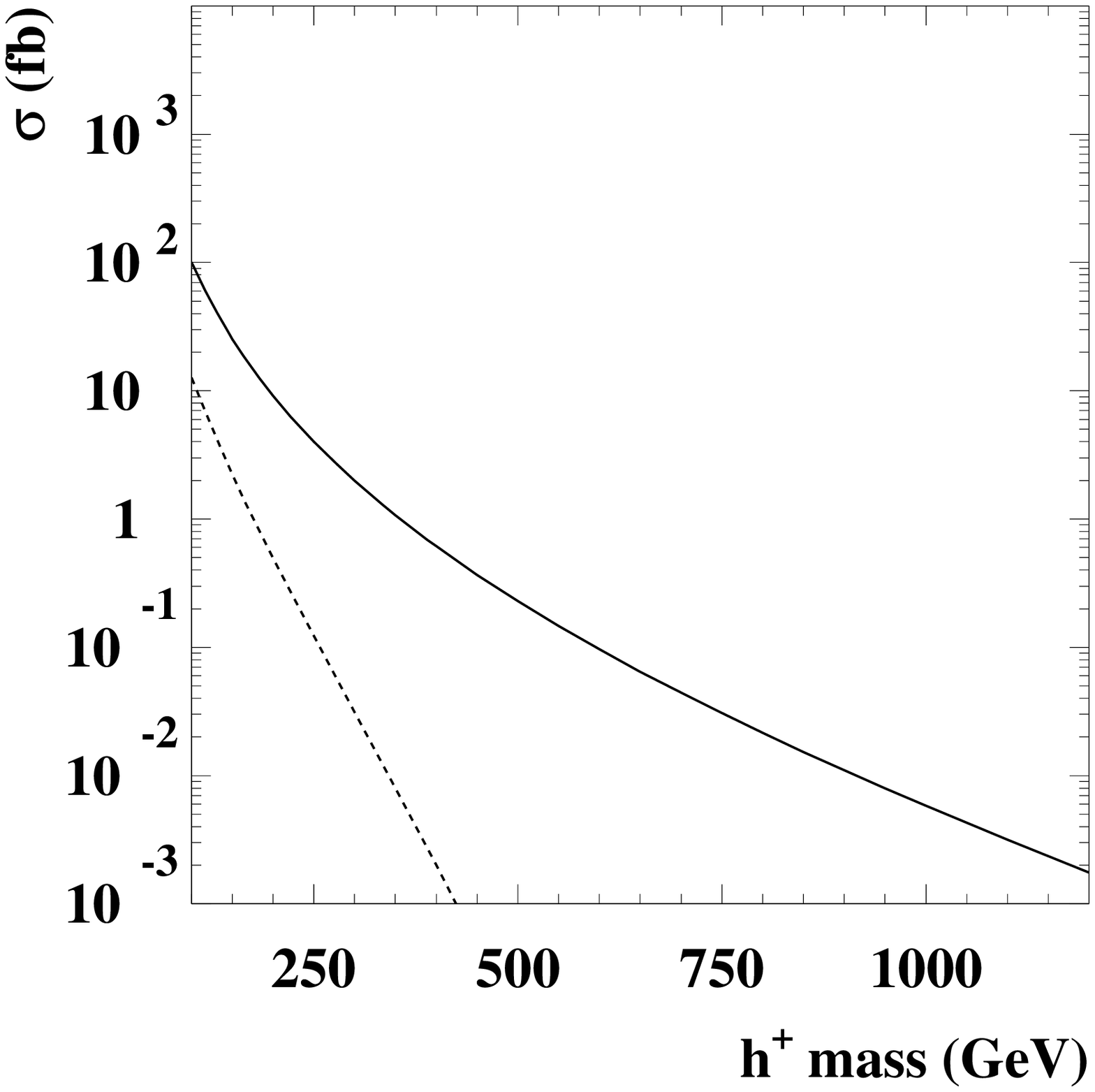}
   \includegraphics[height=3.5in]{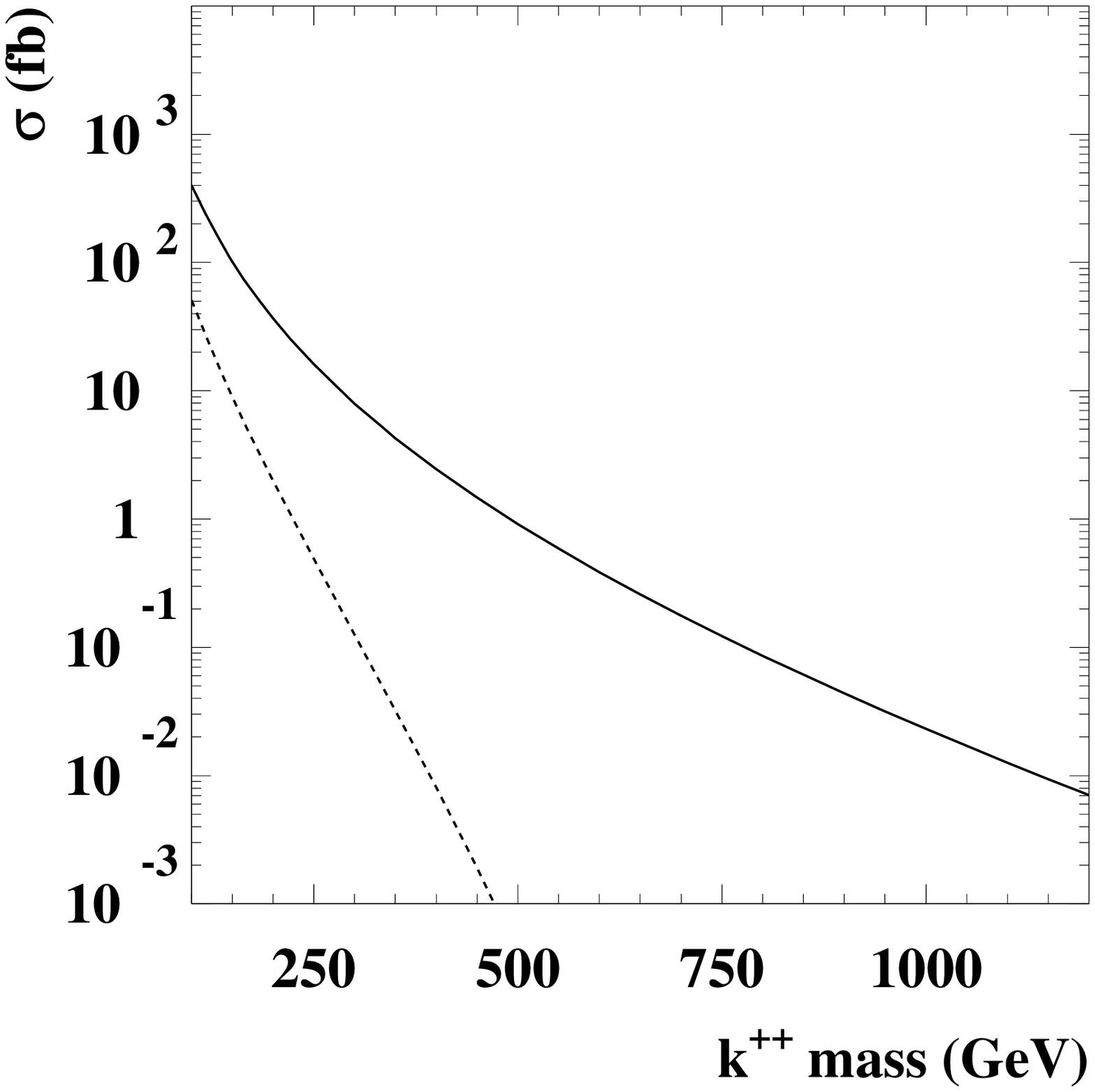}
   }
\caption{Scalar boson production cross-section}
\label{boson_cs}
\end{figure}

\section{Collider Signals}

We now turn to the possible signals coming
from the production of the charged scalars at colliders. Consider first
the case of hadron colliders.  Since the $h^+$ and $k^{++}$ scalars
do not couple
directly to the quarks, the production of these particles will proceed
through the $s$-channel processes: $ q\bar{q} \rightarrow \gamma^* , Z^*
\rightarrow h^+ h^-, k^{++} k^{--}$. The cross-section for these
processes, at the 2 TeV Tevatron and at the LHC, are presented in Fig.
\ref{boson_cs}, as a function of the mass of these particles. As
can be seen from Fig. 4, the cross-section for the production of the
doubly charged scalar $k^{++}$ is about four times larger than the
one for the production of the singly charged scalar $h^+$
(for equal masses). Experimentally,
the $k^{++}$ is also much easier to see; the hierarchy between
the $h$ couplings $h_{\mu \mu} : h_{\mu \tau} : h_{\tau \tau} \ \simeq \
1: m_{\mu}/m_{\tau} : (m_{\mu}/m_{\tau})^2$
(Eqs. (\ref{app_sol1},\ref{app_sol2})) implies that the $k^{++}$
will decay predominantly to a same sign muon pair (or electron pair,
if $h_{ee}$ is of the same order of magnitude as $h_{\mu \mu}$). In any
case, the experimental signature of 4-lepton final state will be striking.
The SM background from $ZZ$ production can be greatly reduced by
imposing high
$p_T$ cuts on the transverse momenta of the leptons,
and by requiring that the invariant mass of opposite sign lepton pairs
be different from the $Z$ mass.
In Ref. \cite{gunion}, it has been estimated that as few as 10 events are
enough for the discovery of a doubly charged boson which decays mostly to
$e $ or $\mu$ pairs. In this case, the Tevatron Run II will be able to probe
up to about 250 GeV with an integrated luminosity of
15 fb$^{-1}$, while the LHC reach will be about
800 GeV with 100 fb$^{-1}$, or 1 TeV with 1 ab$^{-1}$ integrated
luminosity.

At a hadron collider, the $s$-channel production of a $h^+ h^- $ pair
will be much harder to detect experimentally. The final state of two
leptons + missing  energy (associated with the neutrinos coming from
$h^+$ decay) has SM backgrounds coming from $ ZZ,ZW$ (with one lepton
lost) and $WW$ production,
as well as Drell-Yan production of two leptons with mis--measured energy.
$p_T$ cuts can be used to reduce the background in this case
too, but a more detailed
analysis is needed to obtain the hadron collider reach for $h^+$
production in this mode.
Another interesting possibility would be the production of a single $h^+$
through radiation from the $\nu$ line in the process
$ q \bar{q}' \rightarrow l \bar{\nu}_l$. The final state in this case
will be three leptons plus missing $E_T$. High $p_T$ cuts
and cuts on the invariant mass of lepton pairs
can be used here too to reduce the SM
background (which will come mostly from $WZ$ production). However, the
production cross section in this case is proportional to the $f$ couplings
squared. For $ h^+$ boson masses below 1 TeV, the $\mu \rightarrow
e \gamma$ constraint Eq. (\ref{f_cons}) requires that $f$ be in the
0.01 - 0.1 range; as a consequence, there will be at most of order tens
of single $h^+$ events produced at the LHC for any $h^+$ mass in the TeV
range.

At $e^+ e^-$ colliders, $s$-channel pair
production of singly charged or doubly charged
bosons will be limited by the energy of the machine. However,
due to the cleanliness of the environment,
they will be easy to see, if kinematically
accessible. The
reach in mass at these machines will therefore be roughly $\sqrt{s}/2$.
Provided that
the $h_{ee}$ coupling is large, $k^{++}$ can also be singly produced
at an $e^- e^-$ collider. More interesting, though, would be the study
of the  doubly charged scalar at a muon collider,
 since the $h_{\mu \mu}$ coupling
of the  $k^{++}$ has to be large in this model.


\section{Conclusions}

In this paper we have performed a detailed analysis of a specific
two--loop neutrino
mass model. We have shown that the model can accommodate both the
solar and the atmospheric neutrino data, while at the same time satisfying
all bounds arising from leptonic phenomenology.  While at present there
is no conflict with any limit, the model predicts that the rare decays
$\tau \rightarrow 3\mu$ and $\mu \rightarrow e+\gamma$ should be within
reach of forthcoming experiments.

Although the model contains two independent coupling matrices,
one symmetric and the other antisymmetric, they are well
constrained by the current neutrino oscillation data.
We have found a simple way of relating these matrices to neutrino
oscillation parameters.  In
particular, we have shown that the relative ratios between the
three elements of the antisymmetric  coupling matrix $f$ are completely determined by
the neutrino mixing angles, regardless of the other parameters of the model.
Since the contribution of the symmetric couplings $h_{ab}$
to the neutrino mass matrix is proportional to the product
of the masses of the $a$ and $b$ charged
leptons, only $h_{\mu \mu}, h_{\mu \tau}$ and $h_{\tau \tau}$
are relevant for neutrino oscillations. The facts that the atmospheric mixing angle
$\theta_{23}$ is close to $\pi/4$ and that there is a hierarchy in the solar
and atmospheric neutrino
mass--splittings
fixes the relative ratio between  these three parameters to be
$h_{\mu \mu} : h_{\mu \tau} : h_{\tau \tau} \ \simeq \
1: m_{\mu}/m_{\tau} : (m_{\mu}/m_{\tau})^2$.
The fact that  $h_{\mu \mu}$ is the larger of these may have implications
on the search for the doubly charged scalar $k^{++}$ in collider
experiments.

From the experimental result on the atmospheric
neutrino mass splitting and the constraints on the anomalous process
$\mu \rightarrow e \gamma $, a lower bound of 0.01 on the largest of the
$f$ and $h$ coupling constants has been derived. There are upper bounds
on the masses of the charged scalars $h^+$ and $k^{++}$ as well.
For non-maximal values of the Yukawa
couplings (for example, $f_{\mu \tau} \approx
h_{\mu \mu} \approx 0.1$) the predicted
values for the scalar masses are in the TeV range, which should be
probed at the LHC, and perhaps at Run II of the Tevatron.

\section*{Acknowledgments}

This work is supported in part by U.S. Department of Energy grants
DE-FG03-98ER41076 and DE-FG02-01ER45684 and a grant from the Research
Corporation.


\end{document}